# Magnetocaloric effect in the intermetallic compound DyNi

S. K. Tripathy[1], K. G. Suresh[1*], R. Nirmala[2], A. K. Nigam[2] and S. K. Malik[2]

[1]Department of Physics, I.I.T. Bombay, Mumbai - 400076, India
[2]Tata Institute of Fundamental Research, Mumbai - 400005, India

## Abstract

*Magnetic and heat capacity measurements have been carried out on the polycrystalline sample of DyNi which crystallizes in the orthorhombic FeB structure (space group Pnma). This compound is ferromagnetic with a Curie temperature of 59 K. Magnetization-field isotherms at low temperatures shows a step-like behavior characteristic of metamagnetic transitions. The magnetocaloric effect has been measured both in terms of isothermal magnetic entropy change and adiabatic temperature change for various applied magnetic fields. The maximum values of the entropy change and the temperature change are found to be 19 $Jkg^{-1}K^{-1}$ and 4.5 K, respectively, for a field of 60 kOe. The large magnetocaloric effect is attributed to the field-induced spin-flop metamagnetism occurring in this compound, which has a noncollinear magnetic structure at low fields.*

## 1. Introduction

Intermetallic compounds formed between rare earths (R) and transition metals (TM) have attracted considerable attention owing to their potential for various applications[1-3]. Recently, there has been a vigorous activity in the development of magnetocaloric materials, which are the active materials in magnetic refrigerators[4-7]. The magnetocaloric effect (MCE) manifests as the isothermal magnetic entropy change or as the adiabatic

---

*Author to whom all correspondence should be addressed.

Email id: suresh@phy.iitb.ac.in

temperature change of a magnetic material when exposed to a magnetic field. This effect is an intrinsic property of the material and occurs due to the change in the degree of alignment of the magnetic moments in the magnetic sublattice of the system under the influence of an applied magnetic field. The concept of magnetic refrigeration, which is based on MCE, has attracted a great deal of attention from a large group of researchers and has triggered an intensive search for compounds with large MCE. Generally, due to their large magnetic moments, heavy rare earth elements and their compounds are considered the best-suited materials for achieving large MCE. After the observation[1] of a giant MCE in $Gd_5Si_2Ge_2$ and the studies on the pseudo-binary systems[8] like $R_5(Si_xGe_{1-x})_4$ with R= La, Lu, Nd and Dy, much interest has been focused on the compounds showing field-induced magnetic phase transitions and/or structural transitions. But the magnetocaloric studies in $MnFeP_{1-x}As_x$ system[9] has shown that large MCE is not just restricted to the compounds with large magnetic moments but depends strongly on the type of magnetic/structural phase transitions as well. It has already been reported that single crystals of some RNi (R=Dy, Er) compounds undergo field-induced spin flop process, which results in metamagnetic transitions in their magnetization-field isotherms[10]. It would be of interest to see whether such a spin-flop process is reflected in the magnetization behavior in the polycrystalline samples as well and also its influence on the MCE. With this view, we have carried out the magnetic and magnetocaloric studies on the compound DyNi and the results are presented here.

**2. Experimental details**

The polycrystalline sample of DyNi was prepared by arc melting stoichiometric proportions of the starting materials (of at least 99.9% purity) on a water-cooled copper hearth under high purity argon atmosphere. The ingots were melted several times and subsequently annealed at 600°C in evacuated quartz tube to ensure phase formation and homogeneity. The crystal structure and the phase purity of the sample were analyzed from the Rietveld refinement of the powder x-ray diffraction data. Magnetization measurements, in the temperature range 4-150 K and up to a maximum field of 120 kOe, were carried out on pieces of the annealed sample using a vibrating sample magnetometer (VSM, Oxford instruments). Heat capacity measurements, in the temperature range 2-200 K and in magnetic fields up to 50 kOe, were performed using a relaxation method (PPMS, Quantum Design). While the isothermal magnetic entropy change has been calculated independently using the magnetization isotherms as well as the heat capacity data, the adiabatic temperature change has been estimated using the heat capacity data.

**3. Results and discussion**

The Rietveld analysis of the powder x-ray diffractogram of the polycrystalline sample of DyNi showed that this compound is single phase crystallizing in the orthorhombic FeB structure (space group Pnma, no. 62)[11-12]. Figure 1 shows the observed and refined powder x-ray diffractogram of DyNi. The lattice parameters obtained from the refinement are $a = 7.023 \pm 0.001$ Å, $b = 4.173 \pm 0.001$ Å and $c = 5.439 \pm 0.001$ Å which compare well with the reported values[11]. Both Dy and Ni occupy the 4c site of the unit cell. The structure is made up of trigonal prisms with Dy at the corners and Ni at the center.

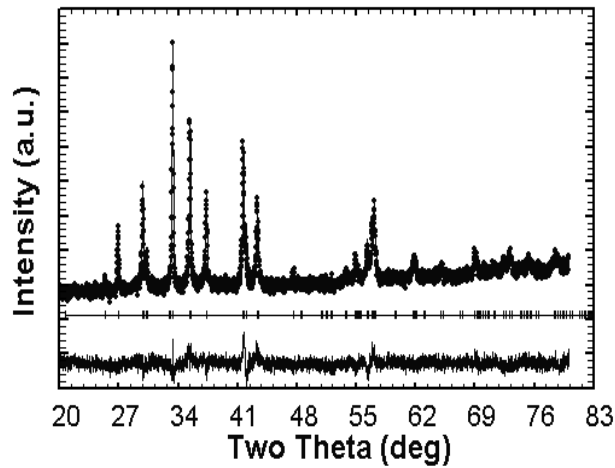

Fig 1. Observed and fitted powder x-ray diffraction patterns of DyNi. The difference plot between the experimental and calculated patterns is given at the bottom of the figure.

Figure 2 shows the temperature dependence of the magnetization (M) of DyNi, measured in a field of 500 Oe. It is seen that the compound is ferromagnetic with a Curie temperature ($T_C$) of 59 K, which is close to the reported value[12]. The inverse susceptibility derived from the magnetization data is presented in the inset of figure 2. The effective magnetic moment ($\mu_{eff}$) calculated from the high temperature susceptibility is found to be 10.6$\mu_B$, which is almost equal to the free ion magnetic moment of 10.3 $\mu_B$ of the $Dy^{3+}$ ion. Figure 3 shows the variation of magnetization for both increasing and decreasing magnetic fields (H) at 4 K. Interestingly, the magnetization shows a multi-step behavior, which is a characteristic of metamagnetic transitions. It is of importance to note that generally such sharp transitions are seen in metamagnetic systems in the single crystalline form. To confirm the existence of these steps, we have measured the M-H at 5 K also and the plot is shown in the inset of figure 3. These plots also show that the magnetization is not saturated even at the highest field applied and the remanence is almost negligible.

It has been reported[10] from the magnetization studies in single crystals of DyNi that when the applied field is along the c-axis, the magnetization is very small at lower fields, but rises sharply above a critical field. This gives rise to a step-like behavior, accompanied by a large hysteresis. The critical field is reported to be about 50 kOe at 4.2 K. Both the critical field and the hysteresis decrease with increase of temperature. Neutron diffraction studies[12] carried out at 4.2 K on powder samples have shown that the magnetic structure of DyNi is non-collinear. While the moments are ferromagnetically coupled along the *a*-axis, they are antiferroimagnetically coupled along the c-axis. The non-collinear magnetic structure of DyNi arises from the compromise between the exchange interaction and the magnetocrystalline anisotropy caused by the crystalline electric fields. The antiferromagnetic component is expected to undergo a spin flopping on application of a magnetic field along the c-axis, which gives rise to a sudden increase in the magnetization. Therefore, we feel that the multi-step behavior of the M-H plot in the polycrystalline DyNi in the present case is also due to the spin-flop metamagnetism. However, while the single crystal data showed a single jump, our data on polycrystalline sample show multiple jumps. A similar behavior in the M-H isotherms has been observed in polycrystalline samples of $Ce_3PbC$, which also possesses a complex magnetic structure[13].

The saturation magnetic moment obtained from the M vs. 1/H plot at 4 K is found to be 8.8 $\mu_B$ per Dy ion whereas the theoretical gJ value of $Dy^{3+}$ ion is 9.75 $\mu_B$. The lower value in the present case may be due to the crystal field effects. Figure 4 shows the magnetization as a function of applied field at different temperatures close to $T_C$. It is clear from figures 3 and 4 that the magnetization behaves like that in a simple ferromagnet and that the remanence is quite negligible, which indicates that the compound is magnetically soft.

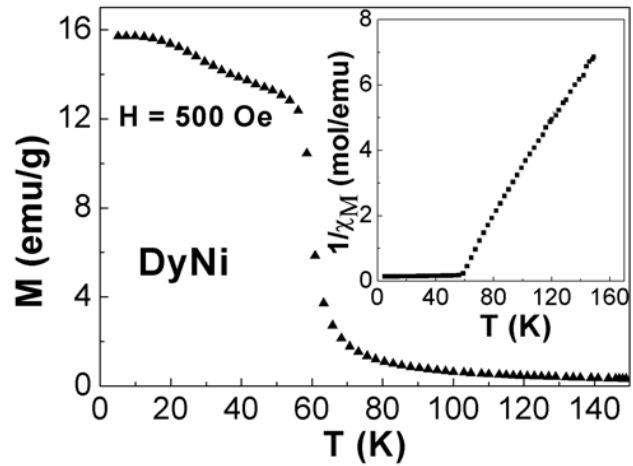

Fig. 2 Temperature (T) dependence of the magnetization (M) of DyNi in a field of 500 Oe. The inset shows the variation of the inverse molar susceptibility with temperature.

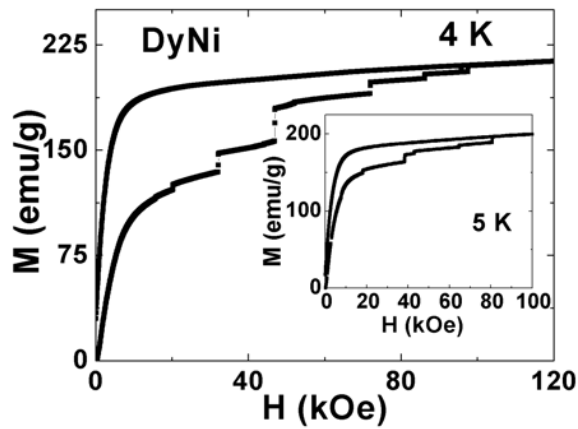

Fig. 3. Magnetization (M) as a function of applied field (H) (increasing and decreasing) for DyNi at 4 K. The inset shows the M-H plot at 5 K.

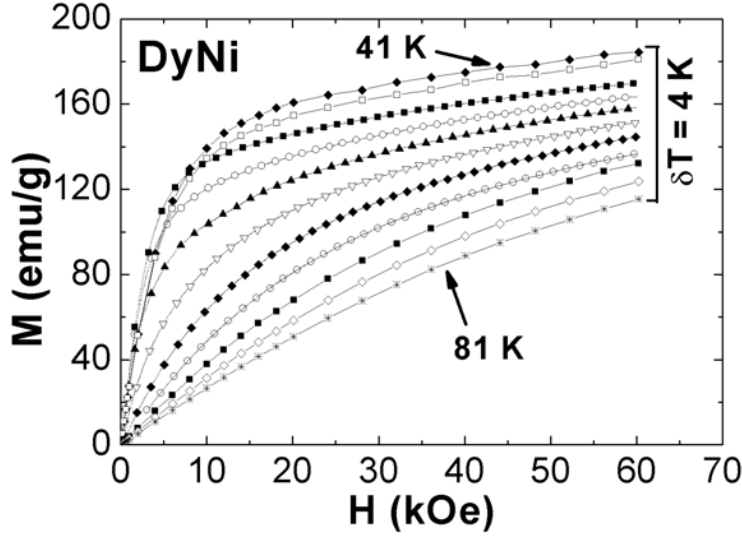

Fig. 4. Magnetization-field (M-H) isotherms for DyNi at different temperatures close to $T_C$.

The MCE in terms of the magnetic entropy change, $\Delta S_M(T, \Delta H)$, has been estimated from the magnetization isotherms, which were measured in the vicinity of $T_C$ with a temperature interval of 4 K, for a maximum field of 60 kOe for each of the isotherms. The $\Delta S_M(T, \Delta H)$ values were calculated by numerically integrating the Maxwell equation[6],

$$\left(\frac{\partial S(T,H)}{\partial H}\right)_T = \left(\frac{\partial M(T,H)}{\partial T}\right)_H \quad (1)$$

i.e. using the approximation,

$$\Delta S_m(T_{av,i}, H_2) = \frac{1}{T_{i+1} - T_i} \int_0^{H_2} (M(T_{i+1}, H) - M(T_i, H)) dH \quad (2)$$

where $T_{av,i}$ is the average of $T_i$ and $T_{i+1}$.

Figure 5 shows the variation of $\Delta S_M$ for DyNi as a function of temperature close to $T_C$, for $\Delta H$ = 20 kOe, 50 kOe and 60 kOe. As can be seen from the figure, the maximum entropy change of about 19 JKg$^{-1}$K$^{-1}$ is obtained at the transition temperature for $\Delta H$ = 60 kOe. Moreover, the magnitude of $\Delta S_M$ is considerably large over a wide temperature range near T$_C$.

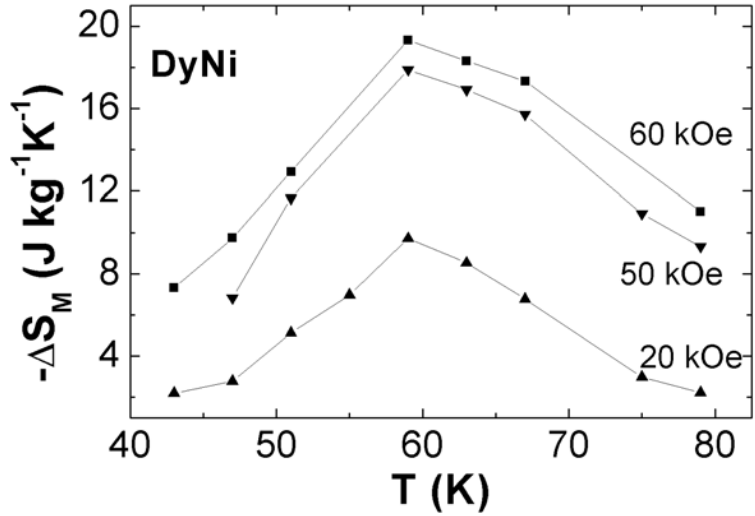

Fig. 5. Variation of the isothermal magnetic entropy change ($\Delta S_M$) as a function of temperature (T) in DyNi for $\Delta H$=40 kOe and 60 kOe.

In order to further understand the magnetocaloric properties of DyNi in more detail, we have carried out heat capacity (C) measurements on this compound under various applied fields. Figure 6 shows the temperature variation of heat capacity in applied fields of 0, 20 kOe and 50 kOe. It is seen from this figure that the heat capacity also shows a peak close to 59 K, which is the Curie temperature derived from the M-T data. The peak height decreases with increase in field, and for a field of 50 kOe, practically, there is no peak in the C-T plot. This is because the magnetic ordering starts at temperatures slightly above T$_C$ when the applied field is large, thereby broadening the heat capacity peak.

8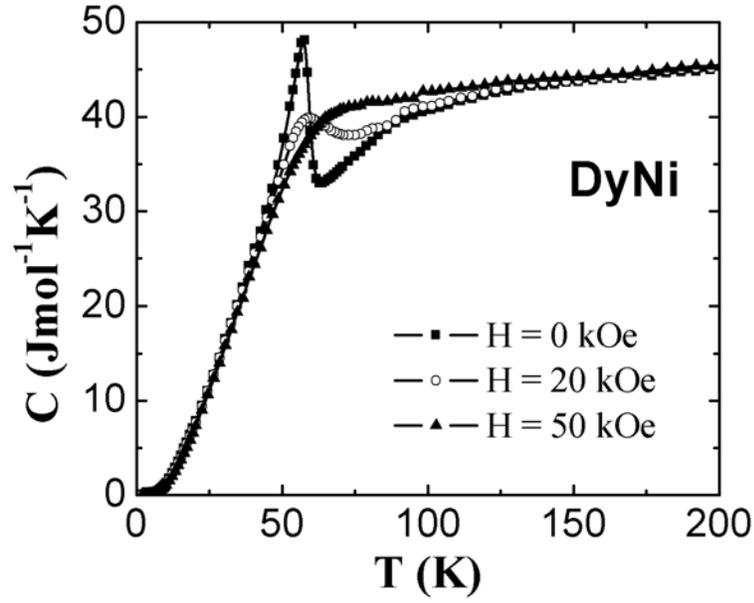

Fig. 6. Heat capacity (C) of DyNi vs. temperature (T) in applied fields of 0, 20 kOe and 50 kOe.

Using the C-H-T data, we have calculated $\Delta S_M$ as well as the adiabatic temperature change ($\Delta T_{ad}$) using the relations[14],

$$\Delta S_M(T,H) = \int_0^T \frac{C(T',H) - C(T',0)}{T'} dT' \qquad (3)$$

$$\Delta T_{ad}(T)_{\Delta H} \cong [T(S)_{H_F} - T(S)_{H_I}] \qquad (4)$$

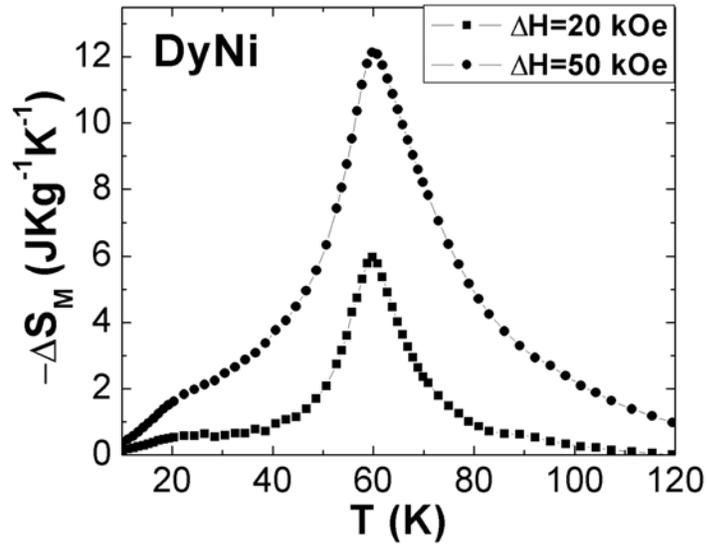

Fig. 7. Variation of magnetic entropy change ($\Delta S_M$) calculated from the heat capacity data as a function of temperature (T) for $\Delta H$ =20 kOe and 50 kOe in DyNi.

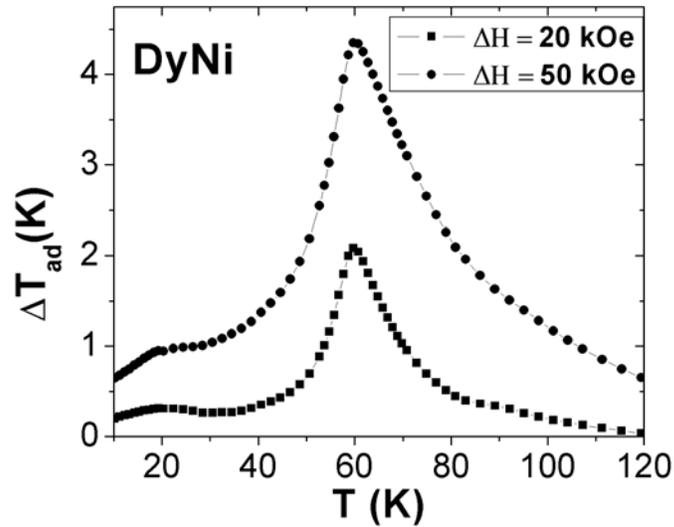

Fig. 8. Adiabatic temperature change ($\Delta T_{ad}$) as a function of temperature (T) for $\Delta H$ =20 kOe and 50 kOe in DyNi.

Figures 7 and 8 show the variation of $\Delta S_M$ and $\Delta T_{ad}$ calculated using the C-H-T data. The maximum value of $\Delta S_M$ obtained from the heat capacity data is found to be about 12.5

Jkg$^{-1}$K$^{-1}$ whereas the corresponding value from the magnetization isotherms is about 18 Jkg$^{-1}$K$^{-1}$ for the same field change of 50 kOe. This difference may be due to the different temperature steps (δT) used for the magnetization and heat capacity measurements. Since the numerical integration depends on δT, the values of Δ$S_M$ calculated from the M-H-T and C-H-T data are also expected to show a difference, as seen in the present case.

It is of interest to note that the Δ$S_M$ and Δ$T_{ad}$ values obtained for DyNi are comparable to those of many potential magnetic refrigerant materials[2,3,15]. Moreover, the temperature range over which the MCE is considerable is also wide, which is essential for a practical magnetic refrigerant. The large MCE values in DyNi can be attributed to its metamagnetic nature mentioned earlier. It is possible that the gradual flopping of moments towards the direction of the applied field is responsible for the reduction in magnetic entropy, which results in MCE. Spin flop process has also been found to be responsible for the anomalous magnetoresistance observed along the c-axis in single crystals of DyNi[16].

## 4. Conclusions

In conclusion, we find that the compound DyNi, in the polycrystalline form, shows a multi-step magnetization behavior, indicating its metamagnetic nature. Large values of isothermal magnetic entropy change and adiabatic temperature change existing over a wide temperature range make this system attractive from the point of view of magnetic refrigeration applications in the temperature range 50-80 K. The fact that the compound is magnetically soft is another advantage of this material.

**Acknowledgement**

One of the authors (KGS) thanks D.S.T., Govt. of India, for financial support in the form of a sponsored project.